# SIMULATION AND ANALYSIS OF LIST MODE MEASUREMENTS ON SILENE REACTOR

**Philippe Humbert**

CEA, DAM, DIF, F-91297 Arpajon, France

philippe.humbert@cea.fr

**ABSTRACT**

Neutron time correlation measurements are used to characterize fissile systems. For this purpose, detailed information are obtained using list mode measurements where the neutron detection times are recorded in a file called the time list file. In this article, we present the simulation and analysis of list mode measurements performed on SILENE reactor. Analogue Monte Carlo simulations are performed using MCNPX code and the synthetic time list file is obtained by processing the MCNPX-PTRAC coincidence capture file. Both measured and computed time list files are processed using the same postprocessor that we have developed. This software is used to calculate counting statistics data like counting number probability distribution, Rossi alpha and Feynman functions for time gates of increasing duration. The parameter uncertainties are evaluated using a block bootstrap method.

## I. INTRODUCTION

The analysis of neutron time correlations is a technique used to characterize fissile materials[1-3] [1-3]. In a fissile assembly, each fission gives rise to a fission chain consisting of all the neutrons that have for ancestor that same initial fission. Neutrons that belong to the same chain stay correlated in time. In practise, the most complete correlation measurements are done in list mode, which consists of recording in a file the list of the counting times or the time intervals between successive neutron detections. There are different methods for extracting useful information from time list files. Here, we are interested in the counting statistics during time windows of increasing duration. Thus, we obtain the counting number probability distributions and their first moments.

In the following, we present the simulation and analysis of list mode measurements performed with the SILENE reactor [4]. The experiment is simulated using MCNPX Monte Carlo code [5] in analog mode. The result of the simulation is a coincidence capture file, the PTRAC file [6] that is used to generate a synthetic time list file. The calculated and measured time list files are analysed using the same postprocessor that we have developed. The time list file analysis is based on the first three correlated moments of the counting distribution. The second order moment method was first proposed by Feynman [7] and has been extended to the third order by different authors [8-11]. The interest of the method is that within the point reactor approximation closed expressions have been derived that relates the measurements to the fissile system and detector characteristic parameters (reactivity, time constant, source intensity, spontaneous fission fraction, detector efficiency).

## II. SECOND AND THIRD ORDER FEYNMAN FUNCTIONS

### II.A. Detection of correlated neutrons

The statistical analysis of the number of detected neutrons in a time window of duration $T$ is used to evaluate the characteristic parameters of fissile assemblies. The measured quantities of interest are:

- $\langle C(T)\rangle$ : The expected value of the detection number during a time gate of width $T$,
- $\langle C_2(T)\rangle$ : The expected value of double correlated detections, i.e. the average number of combinations of two correlated detections (cf. fig. 1),
- $\langle C_3(T)\rangle$ : The expected value of triple correlated detections, i.e. the average number of combinations of three correlated detections (cf. fig. 2).

Detected neutrons are correlated when they belong to the same fission chain. The correlated doubles and triples are related to the Feynman second and third order factors.

$$Y(T) = \frac{2\langle C_2\rangle}{\langle C\rangle} \tag{1}$$

$$X(T) = \frac{6\langle C_3\rangle}{\langle C\rangle} \tag{2}$$

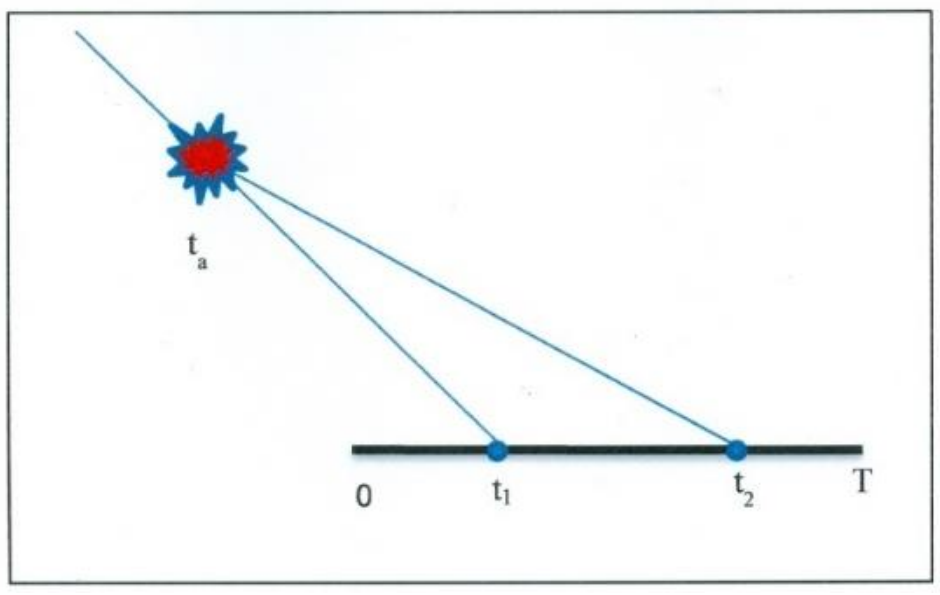

***Figure 1.*** Double correlated detection: Two neutrons that belong to the same fission chain are detected at time $t_1$ and $t_2$ in a time gate of duration $T$. The lines after the initial fission represent fission chains.

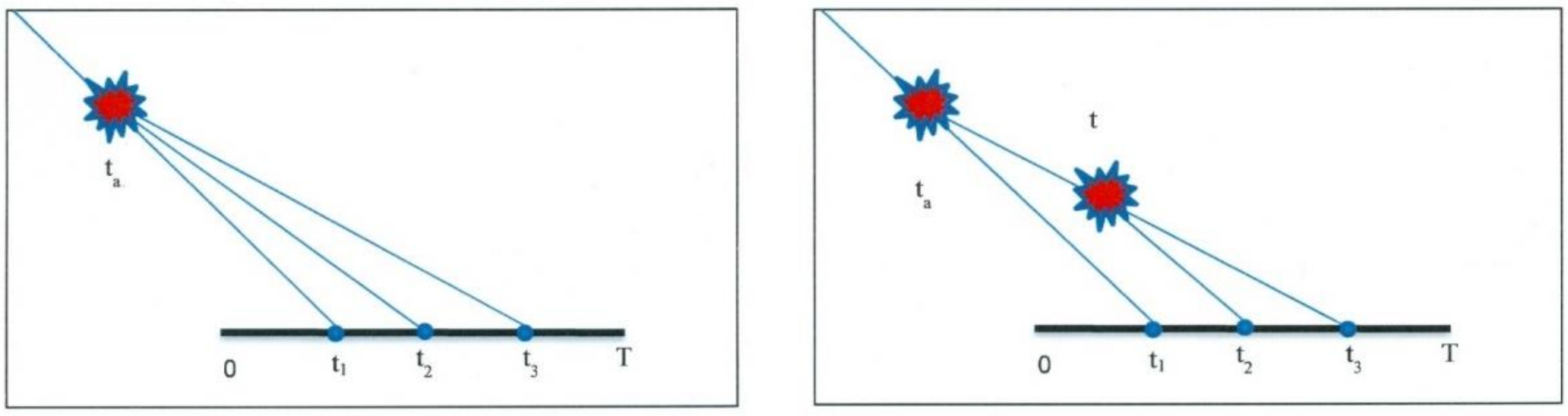

***Figure 2.*** Two configurations for the triple correlated detection. The lines after the initial and intermediate fission represent fission chains.

The Feynman factors can be expressed in terms of the mean $\mu$ and central moments $\mu_n$ ($n > 1$) of the number of detected neutrons $C(T)$. Using the bracket notation for the expectation, we have:

$$\mu = \langle C \rangle \tag{3}$$

$$\mu_n = \langle (C - \langle C \rangle)^n \rangle \,, \qquad n > 1 \tag{4}$$

The second and third Feynman factors write:

$$Y = \frac{\mu_2}{\mu} - 1 \tag{5}$$

$$X = \left(\frac{\mu_3}{\mu} - 1\right) - 3\left(\frac{\mu_2}{\mu} - 1\right) \tag{6}$$

The Feynman factors are equal to zero for a Poisson counting process when the detected signals are uncorrelated which is the case in the absence of fission.

Let $N$ be the number of counting gates, and $c_i$ the number of counts in the $i^{\text{th}}$ gate, then the moment estimators are as follows:

$$\bar{\mu} = \frac{1}{N}\sum_{i=1}^{N} c_i \tag{7}$$

$$\bar{\mu}_2 = \frac{1}{N-1}\sum_{i=1}^{N} (c_i - \bar{\mu})^2 \tag{8}$$

$$\bar{\mu}_3 = \frac{N}{(N-1)(N-2)}\sum_{i=1}^{N} (c_i - \bar{\mu})^3 \tag{9}$$

The second and third central moment estimators can be considered as unbiased as long as the correlations between gates is negligible, that is when $\langle c_i c_j \rangle \cong \mu^2$. The estimated Feynman factors are:

$$\bar{Y} = \frac{\bar{\mu}_2}{\bar{\mu}} - 1 \tag{10}$$

$$\bar{X} = \left(\frac{\bar{\mu}_3}{\bar{\mu}} - 1\right) - 3\left(\frac{\bar{\mu}_2}{\bar{\mu}} - 1\right) \tag{11}$$

**II.B. Point reactor model**

Within the framework of the point model characterized by mono kinetic neutrons moving in an infinite homogeneous medium, the average number of detections and Feynman factors are expressed in closed form in terms of the system and detector characteristic parameters (cf. the appendix in [12]).

We use the following notations:

| | |
|---|---|
| $\varepsilon_F$ | Detector efficiency defined as the number of counts per induced fission, |
| $Q$ | source intensity (neutrons/s) |
| $\rho$ | Prompt reactivity |
| $\alpha$ | Prompt neutron time constant ($s^{-1}$) |
| $\tau$ | Time gate length (s) |
| $\bar{\nu}$ | Average number of neutrons produced per induced fission |
| $\bar{\nu}_s$ | Average number of neutrons produced per spontaneous fission |
| $x$ | Ratio of spontaneous fission source neutrons to the total source neutrons |
| $D_2 = \frac{\overline{\nu(\nu-1)}}{\bar{\nu}^2}$ | Induced fission Diven factor |
| $D_3 = \frac{\overline{\nu(\nu-1)(\nu-2)}}{\bar{\nu}^3}$ | Third order induced fission Diven factor |
| $D_{2s} = \frac{\overline{\nu_s(\nu_s-1)}}{\bar{\nu}_s^2}$ | Spontaneous fission Diven factor |
| $D_{3s} = \frac{\overline{\nu_s(\nu_s-1)(\nu_s-2)}}{\bar{\nu}_s^3}$ | Third order spontaneous fission Diven factor |

The average number of detections is given by equation (12)

$$\langle C(T)\rangle = -\frac{\varepsilon_F Q}{\rho\bar{\nu}}T \tag{12}$$

The second order Feynman factor is given by equations (13-15). It reaches an asymptotic value $Y_\infty$ when the time gate length is large.

$$Y(\alpha T) = Y_\infty f_Y(\alpha T) \tag{13}$$

$$Y_\infty = \frac{\varepsilon_F D_2}{\rho^2}\left(1 - x\rho\frac{\bar{\nu}_s D_{2s}}{\bar{\nu} D_2}\right) \tag{14}$$

$$f_Y(\alpha T) = 1 - \frac{1 - e^{-\alpha T}}{\alpha T} \tag{15}$$

The third order Feynman factor is given by equations (16-20). It has two terms that correspond to the two detection configurations of figure 2.
It also reaches an asymptotic value $X_\infty = X_{2\infty} + X_{3\infty}$ when the time gate length is large.

$$X(\alpha T) = X_{2\infty} f_{X2}(\alpha T) + X_{3\infty} f_{X3}(\alpha T) \tag{16}$$

$$X_{2\infty} = 3\left(\frac{\varepsilon_F D_2}{\rho^2}\right)^2\left(1 - x\rho\frac{\bar{\nu}_s D_{2s}}{\bar{\nu} D_2}\right) \tag{17}$$

$$X_{3\infty} = -\frac{\varepsilon_F^2 D_3}{\rho^3}\left(1 - x\rho\frac{\bar{\nu}_s^2 D_{3s}}{\bar{\nu}^2 D_3}\right) \tag{18}$$

$$f_{X2}(\alpha T) = 1 - e^{-\alpha T} - 2\frac{1 - e^{-\alpha T}}{\alpha T} \tag{19}$$

$$f_{X3}(\alpha\tau) = 1 - \frac{3 - 4e^{-\alpha T} + e^{-2\alpha T}}{2\alpha T} \tag{20}$$

## III. SILENE REACTOR EXPERIMENT

### III.A. SILENE Reactor

The SILENE facility is an experimental reactor that was put into operation on the CEA/Valduc site until it was shut down in 2010 [13]. This reactor was initially designed to study the phenomenology of criticality accidents in fissile solutions. Subsequently, it has also been used as a radiation source for safety and radiation protection studies. It has several modes of operation, the burst mode (supercritical with a short and intense power excursion), the free evolution mode and the stationary mode (subcritical).

The reactor is of small dimensions (cf. figure 3). It consists of a cylindrical vessel of 36 cm internal diameter with à 7 cm diameter central channel in which can be inserted detectors or samples to be irradiated. The tank contains the liquid fuel of the reactor. It is a fissile solution of uranyl nitrate, highly enriched in $^{235}$U (93.5% in weight) and with a uranium concentration of 71g/l. The reactivity of the system depends on the height of the solution.

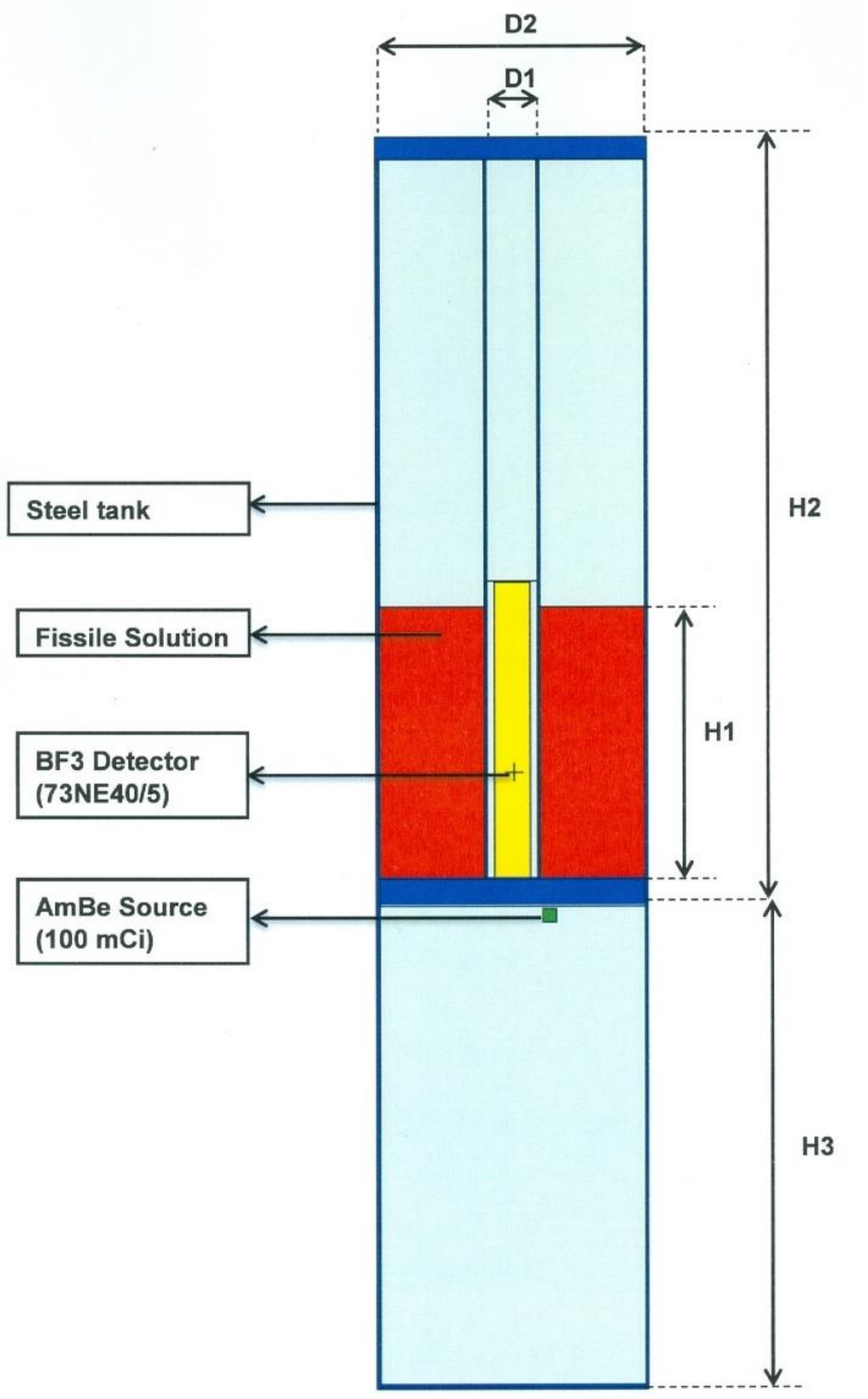

***Figure 3.*** SILENE reactor with fissile solution, BF3 detector and AmBe source.

### III.B. Experiment

The purpose of the experiment was to study neutron noise methods used to characterize the reactivity. The reactor was operated in subcritical mode with a solution height of 30 cm. The emitted neutrons were detected by means of a BF3 detector disposed in the central channel of the reactor, in the centre of the fissile solution. A 100 mCi AmBe source was placed under the fissile solution in order to increase the counting rate. The list mode data acquisition system made it possible to record the list of time intervals between successive neutron detections (time list file). The measurement duration was 663.55 seconds and 1 292 123 detection times were recorded.

## IV. TIME LIST FILE PROCESSING

### IV.A. Sequential Binning

The time list file contains the list of the time delays between neutron detections (cf. figure 4).

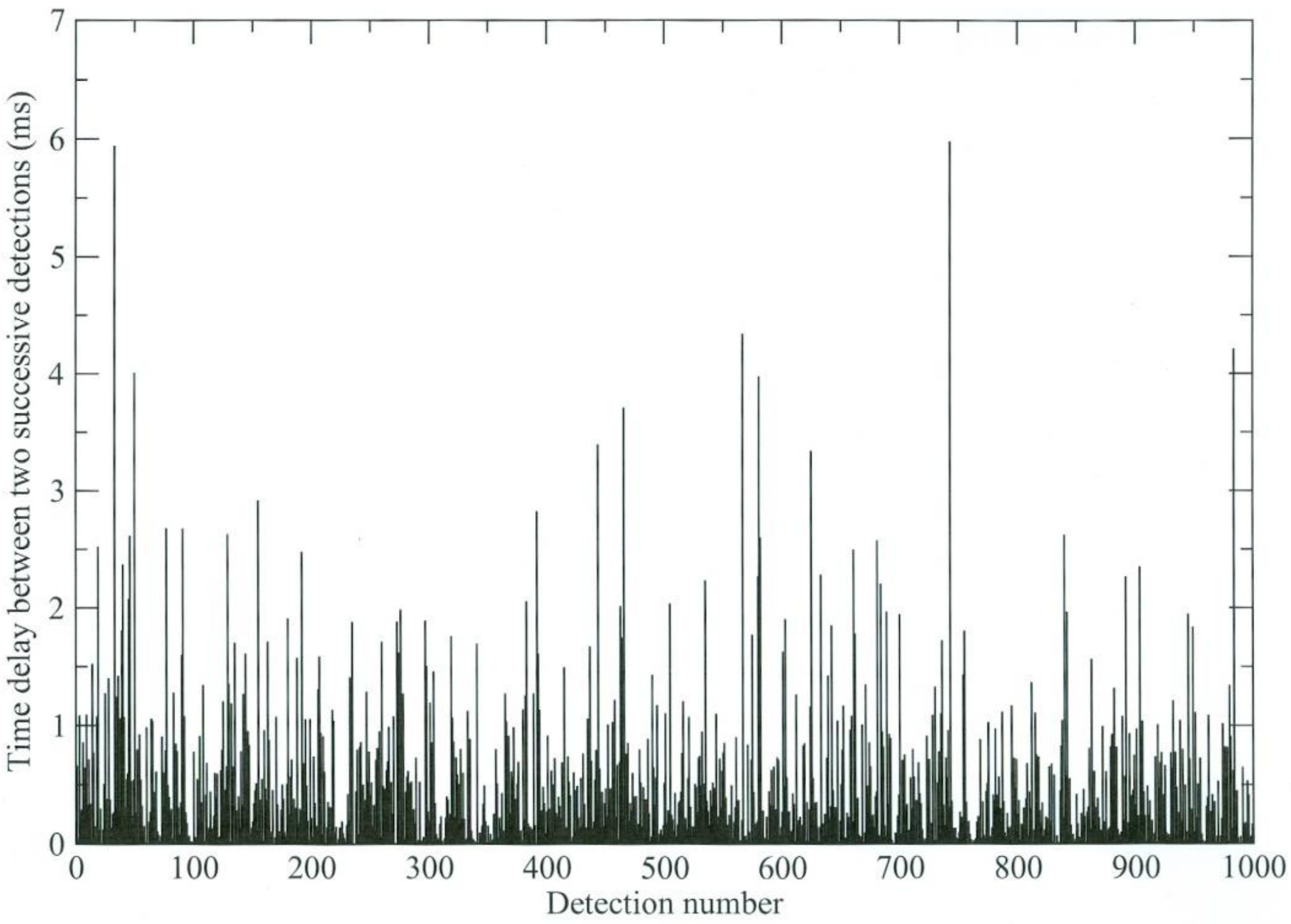

***Figure 4.*** Representation of the time list file that contains the list of time intervals between signals.

To extract the Feynman factors from the time list file, the recorded pulse train is successively split into time intervals of increasing duration. For this purpose, it is divided into a series of consecutive contiguous time gates of length T. Coarser gates of size 2T, 3T … are obtained by collapsing the fine gates of length T (cf. figure 5).

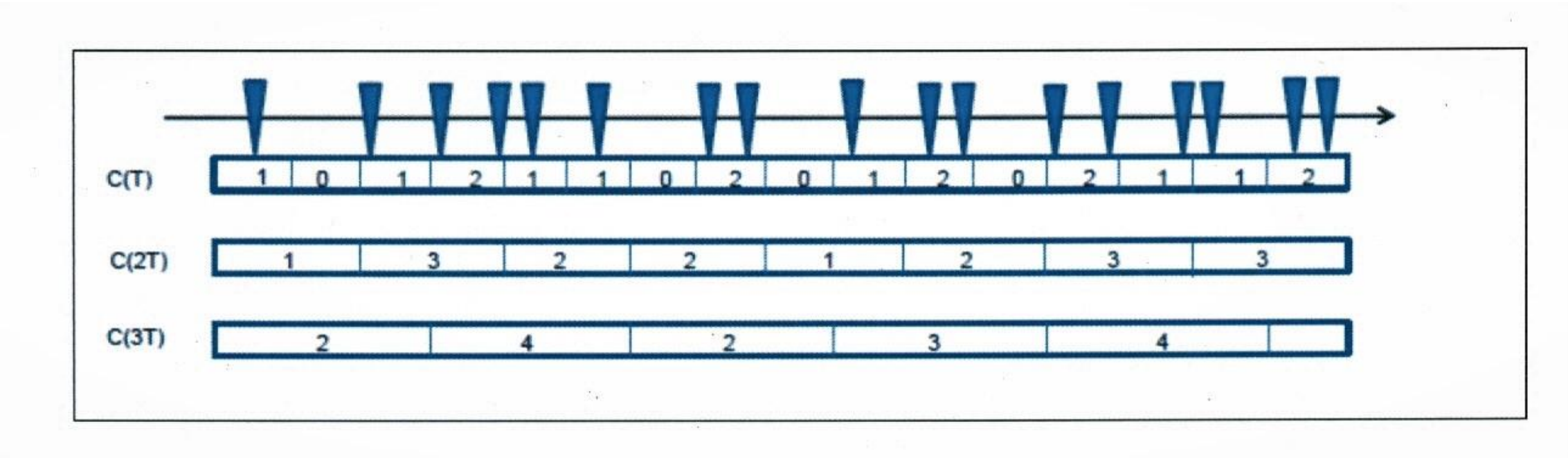

***Figure 5.*** Sequential binning: The pulse train is successively sampled into contiguous gates that are then collapsed into larger gates.

### IV.B. Feynman Parameters Estimation Using Bootstrap Method

The Feynman parameters $(Y_\infty, X_\infty, \alpha)$ are obtained using the Levenberg-Marquardt (LM) error weighted non-linear least square regressions [14] of the measured Feynman factors (eq. 10,11) using the point model expressions (eq. 13-20).

The estimation of the parameter uncertainties is not straightforward because the points of the Feynman's curves are correlated. The bootstrap method [15,16] is used for this purpose. The bootstrap method is a resampling method for estimating the distribution of a statistical parameter. The statistics are made on a set of replicas of the initial sample. The replicas (bootstrap samples) are produced by random draw with replacement. The replicas have the same size as the original sample. In principle, this method requires that the data are independent and identically distributed (iid). When the data are correlated as is the case in time series such as the time list file we can use a block bootstrap method. The resampling is done by blocks instead of individual data. The block size must be sufficient to keep the original structure within each block. In our case, the time list file has been processed using a non-overlapping block bootstrap (NBB) method [17]. Each block does not contain the same number of detections but correspond to the same counting time. For instance one considers the second order Feynman curve. The time list file is split into $N$ blocks and $B$ bootstrap samples are produced by block resampling with replacement. $Y(b,T)$ is the excess of relative variance calculated for a gate of length $T$ on the time list bootstrap sample number b. The bootstrap mean and standard deviation are:

$$\bar{Y}^B(T) = \frac{1}{B}\sum_{b=1}^{B} Y(b,T) \tag{21}$$

$$\sigma_Y^B(T) = \sqrt{\frac{1}{B-1}\sum_{b=1}^{B}[Y(b,T) - \bar{Y}^B(b,T)]^2} \tag{22}$$

The standard deviations are used as weights in the LM regression. Parameters uncertainties are estimated by performing a regression on each of the $B$ samples. Considering the asymptotic value of the excess of reduced variance, the bootstrap mean and standard deviation are:

$$\bar{Y}_\infty^B = \frac{1}{B}\sum_{b=1}^{B} Y_\infty(b) \tag{23}$$

$$\sigma_{Y_\infty}^B = \sqrt{\tfrac{1}{B-1}\textstyle\sum_{b=1}^{B}[Y_\infty(b) - \bar{Y}_\infty^B]^2} \tag{24}$$

The second and third order Feynman's curves obtained by processing the SILENE experiment time list file is presented in figure 6 which show that the Feynman factors reach an asymptote for large time gates. The experimental results are summarized in table I. The time list file is split into $N$=30 blocks and the number of bootstrap samples is $B$=400. The alpha value is calculated using the $Y(T)$ curve, The alpha value obtained with the $X(T)$ curve is $\alpha_X = 2198$ s$^{-1}$ ($\sigma = 75$ s$^{-1}$).

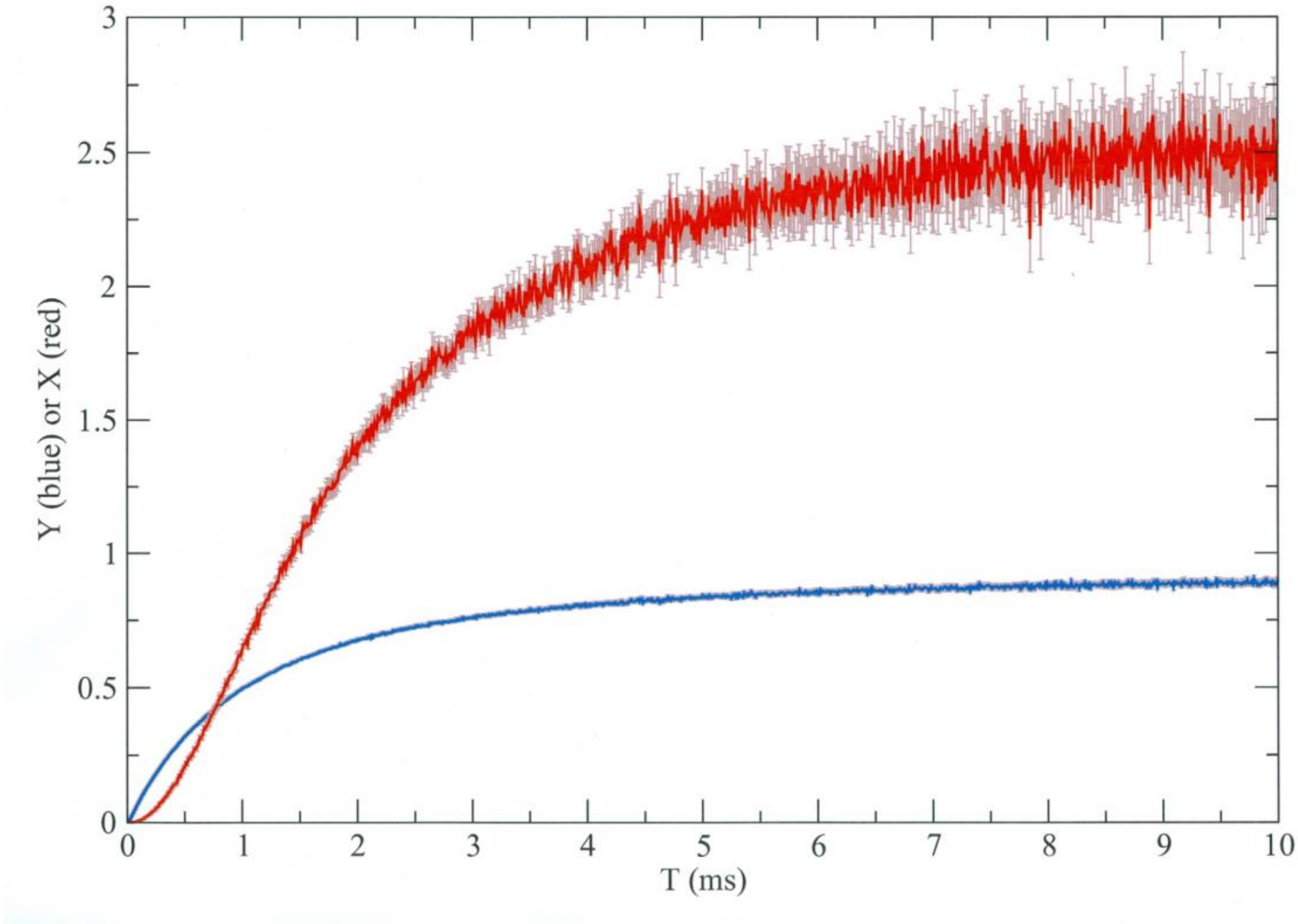

***Figure 6.*** Measured second (blue) and third (red) order Feynman factors in the SILENE experiment. The uncertainties are estimated as the NBB bootstrap standard deviation.

**Table I:** Results of the experimental SILENE time list file processing. The relative uncertainties (standard deviation to the mean) are estimated by the NBB method.

| $C$ ($s^{-1}$) | $\alpha$ ($s^{-1}$) | $Y_\infty$ | $X_\infty$ |
|---|---|---|---|
| 1947<br>(0.2%) | 1756<br>(1.1%) | 0.9420<br>(0.9%) | 2.831<br>(3.8%) |

## V. SIMULATIONS

### V.A. Synthetic Time List File

MCNPX code used in analogue mode, without variance reduction, generates a coincidence capture file, the PTRAC file from which a synthetic time list file can be produced. The PTRAC file contains, for each detected neutron, the initial source event number and the time interval between this event and the capture in the detector medium. As source events occur according to a Poisson process, the time list file is obtained by adding a random time sampled with a uniform law to the capture times of the PTRAC file. The detections that occur after the end of the measurement are brought back to the beginning in order to correct the transitional period, which precedes the stationary state. We then obtain the list of counting times with the associated history (source event) numbers. The list is sorted according to the increasing detection times and the time intervals between successive detections are calculated to produce the time list file. Compared to the experimental files, the synthetic time list files contain the source event numbers. As we will see this additional information is used to reduce the fluctuations in the calculated Feynman factors.

### V.B. Triggered Binning

The experimental time list file was processed using sequential binning, the calculated file is processed using triggered binning [18,19] in order to take advantage of the history number information. Triggered binning is usually used in the Rossi-alpha method [20]. This method consists in determining the counting rate in windows whose opening is triggered by an initial detection. The counting rate is then the sum of two components. The "real" one contains only the detected neutrons, which belong to the same history as the triggering neutron and the "accidental" one that contains the signals that are not correlated to the trigger. In the triggered gates method (cf. figure 7) each detection triggers the opening of a series of $K$ consecutive windows of same duration.

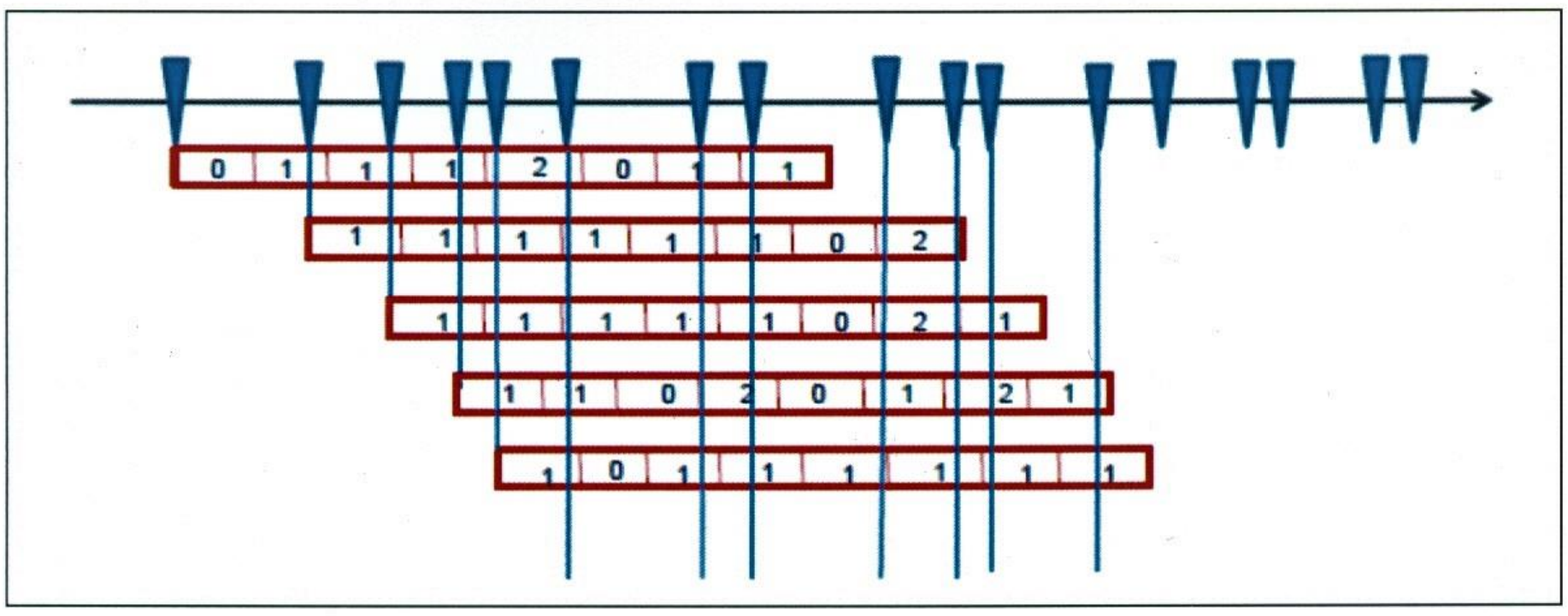

***Figure 7.*** Triggered binning: Each detection triggers the opening of a series of $K$ consecutive windows of same duration.

Using the history number information present in the calculated time list file it is possible to take into account only the "real" component. In the point model approximation, the Rossi-alpha function $R_\alpha(T)$ is the sum of the Accidental ($A$) Real ($R_1$) components given by:

$$R_\alpha(T) = A + R_1(T) \tag{25}$$

$$A = -\frac{\varepsilon_F Q}{\rho \bar{\nu}} \tag{26}$$

$$R_1(T) = \frac{\alpha Y_\infty}{2} e^{-\alpha T} \tag{27}$$

The PTRAC file is used to generate the synthetic time file, which is then processed with the triggered gates method. Examples of Rossi-alpha curves are presented in figure 8 and 9, the curve in figure 8 contains both components (real and accidental) and the curve in figure 9 has only the real component. These figures are obtained from an MCNPX simulation of the SILENE experiment.

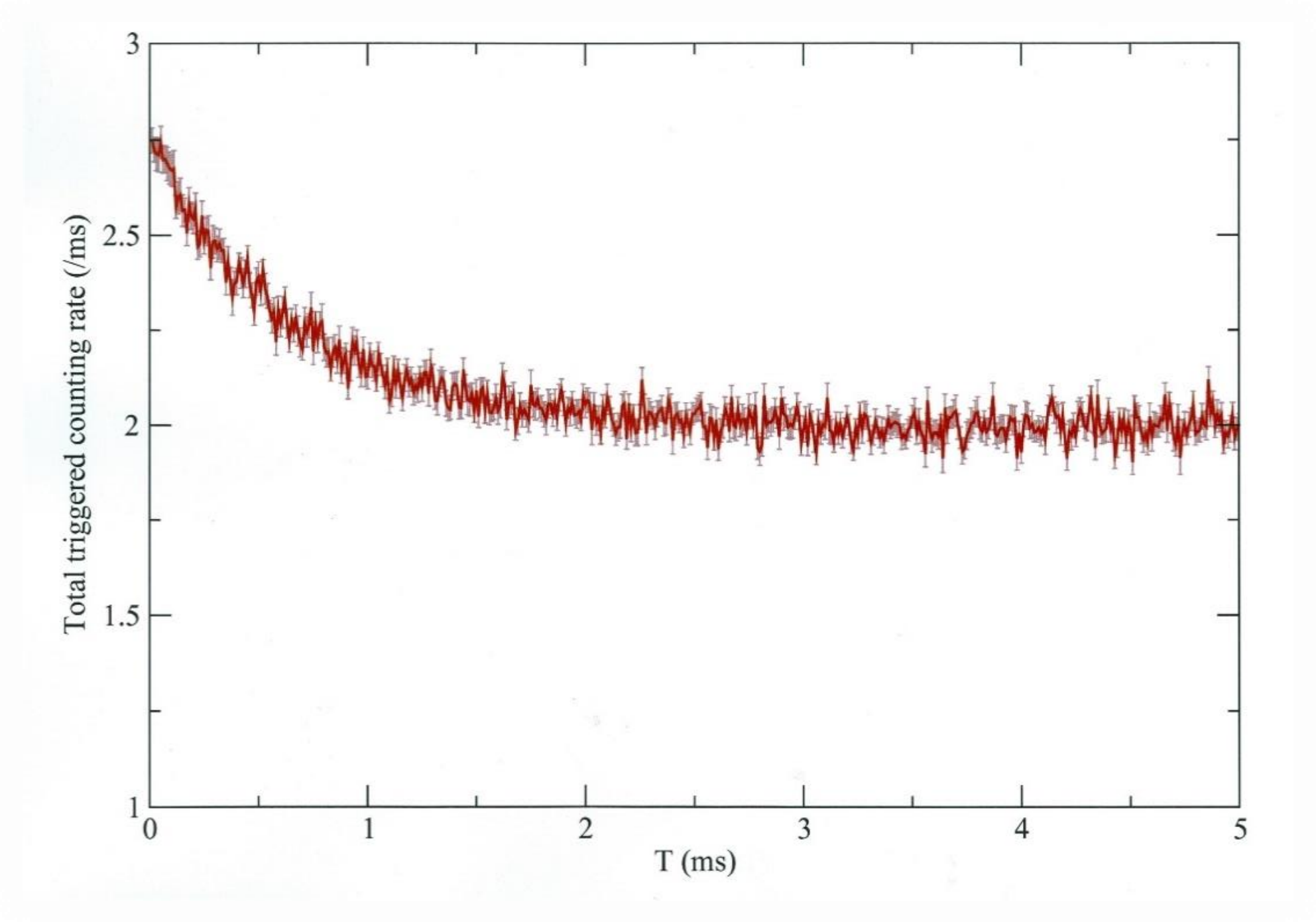

***Figure 8.*** The Rossi alpha curve is the sum of the accidental and real component

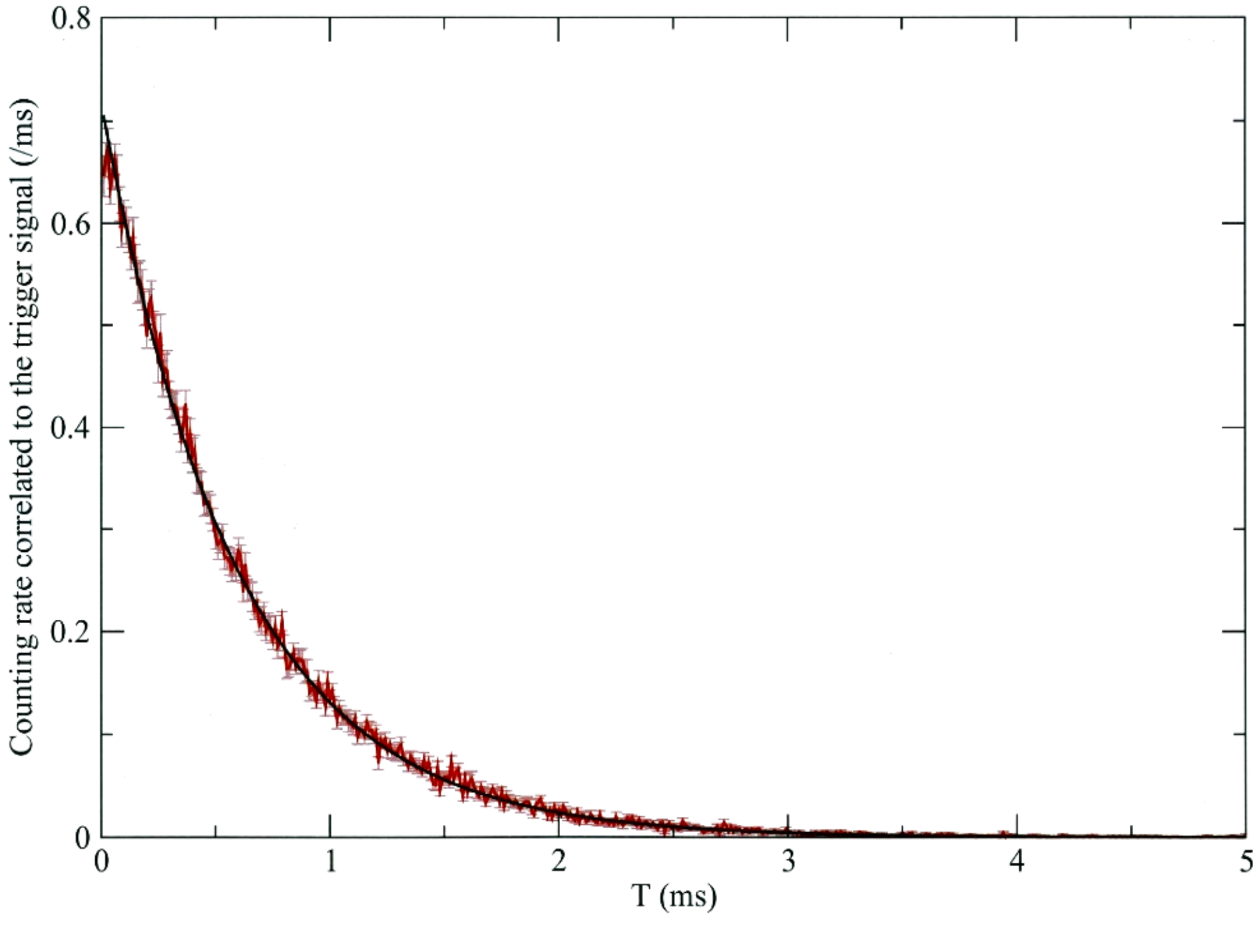

***Figure 9.*** Using the history number available in calculated time list file, the accidental component of the Rossi-alpha curve is removed.

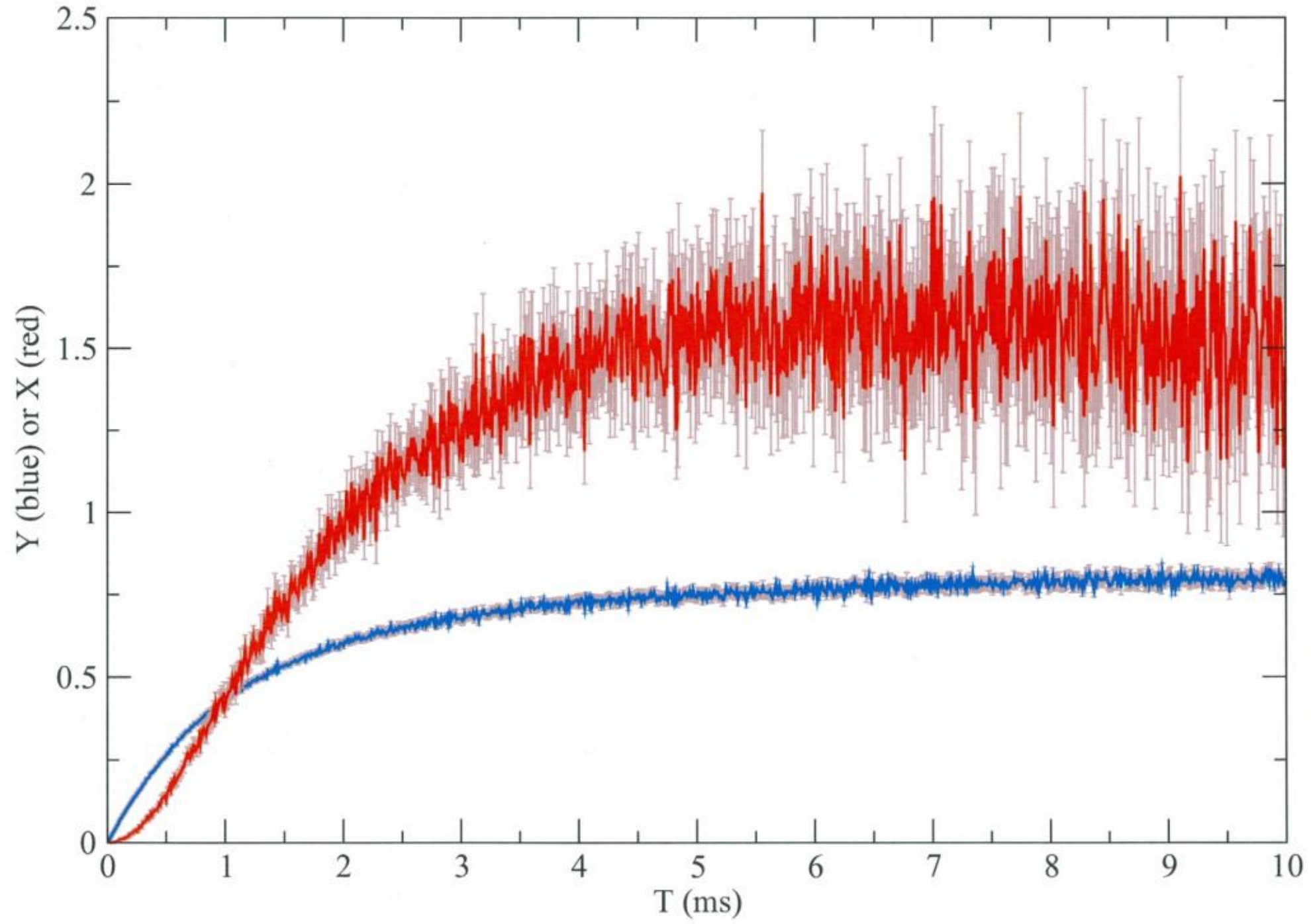

***Figure 10.*** Second (blue) and third order (red) Feynman curves obtained from sequential binning.

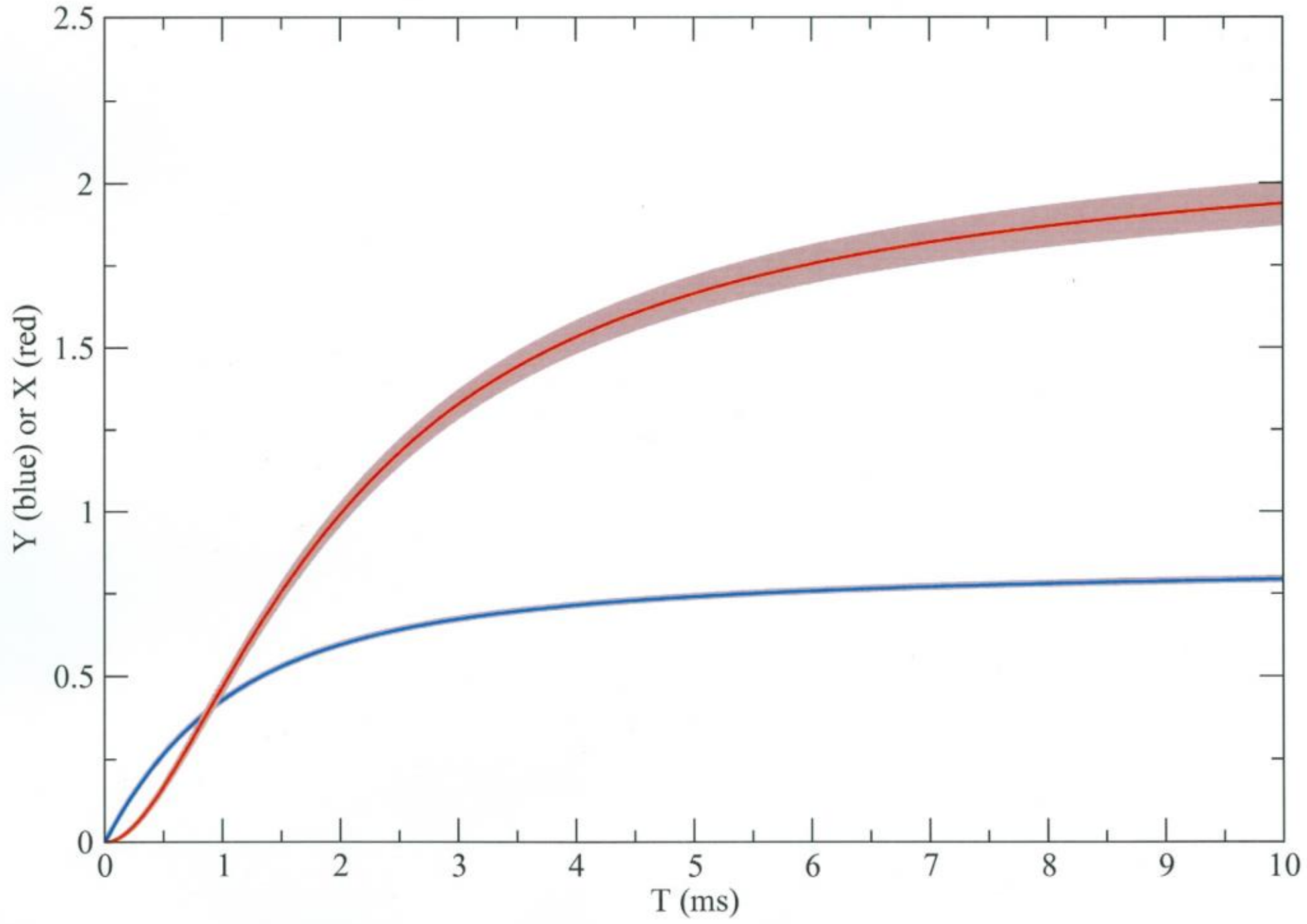

***Figure 11.*** Second (blue) and third order (red) Feynman curves obtained from real counts with triggered binning.

The Feynman's factors are calculated using equations (28-29) where $R_1$ and $R_2$ are the average number of single and double real counts in a triggered gate.

$$Y(T) = \frac{2}{T}\int_0^T R_1(\tau)d\tau \tag{28}$$

$$X(T) = \frac{6}{T}\int_0^T R_2(\tau)d\tau \tag{29}$$

Examples of Feynman curves are presented in figure 10 and 11, the curves in figure 10 are obtained from sequential binning and the curves in figure 11 are calculated using real counts with triggered binning. Both figures are calculated using the PTRAC file of an MCNPX simulation of the SILENE experiment.

### V.C. Results

Monte Carlo simulations of SILENE experiment are performed using MCNPX code with ENDF/B-VII nuclear data [21]. The PTRAC files calculations are made using 27 million source particles corresponding to a measurement time of 100 s. These files are used to generate synthetic time list files that are processed using triggered gates without accidental counts. The uncertainties on the Feynman parameters are determined using the non-overlapping block bootstrap method with 30 blocks and 400 bootstrap samples. The characteristics of the fissile solution used as reference in the simulations are given in [22,23]. This mean composition may be subject to variations during the experiments, reprocessing and various adjustments. The given reference density of the fissile solution is 1.161g/cm$^3$ with an uncertainty evaluated to $\sigma$=5.456 $10^{-3}$ (cf. the appendix I in ref. [23]). Simulations performed with this density underestimate the value of the Feynman parameters $Y_\infty$ and $X_\infty$ which are very sensitive to the system reactivity. A better agreement is obtained with a slightly higher density, i.e. 1.165 g/cm$^3$ within the uncertainty. MCNPX calculations show that this corresponds to an increase of the prompt effective multiplication factor of 239 $10^{-5}$. It is also noted that in contrary, the $\alpha$ value agreed better when using the nominal density. Simulations and measurements results are presented in table II. The relative uncertainties (standard deviation to the mean) are estimated by the NBB method. The calculated and measured Feynman factors $Y(T)$ and $X(T)$ are given in figure 12 and 13 respectively.

**Table II:** Parameters of the Feynman analysis for the experimental and simulated time list files. The simulations are preformed using MCNPX PTRAC files for two values of the fissile density. The relative uncertainties (standard deviation to the mean) are estimated with the bootstrap (NBB) method.

| | MCNPX Simulations | | Experiment |
|---|---|---|---|
| $\rho$ (g/cm$^3$) | 1.161 | 1.165 | |
| $k_p$ | 0.94536 ($\sigma$=0.00013) | 0.94775 ($\sigma$=0.00013) | |
| $C$ (s$^{-1}$) | 1885 (0.4%) | 2001 (0.4%) | 1947 (0.6%) |
| $\alpha$ (s$^{-1}$) | 1625 (1%) | 1585 (0.8%) | 1756 (1.1%) |
| $Y_\infty$ | 0.848 (1%) | 0.948 (1%) | 0.942 (0.9%) |
| $X_\infty$ | 2.203 (3.3%) | 2.758 (3.3%) | 2.831 (3.8%) |

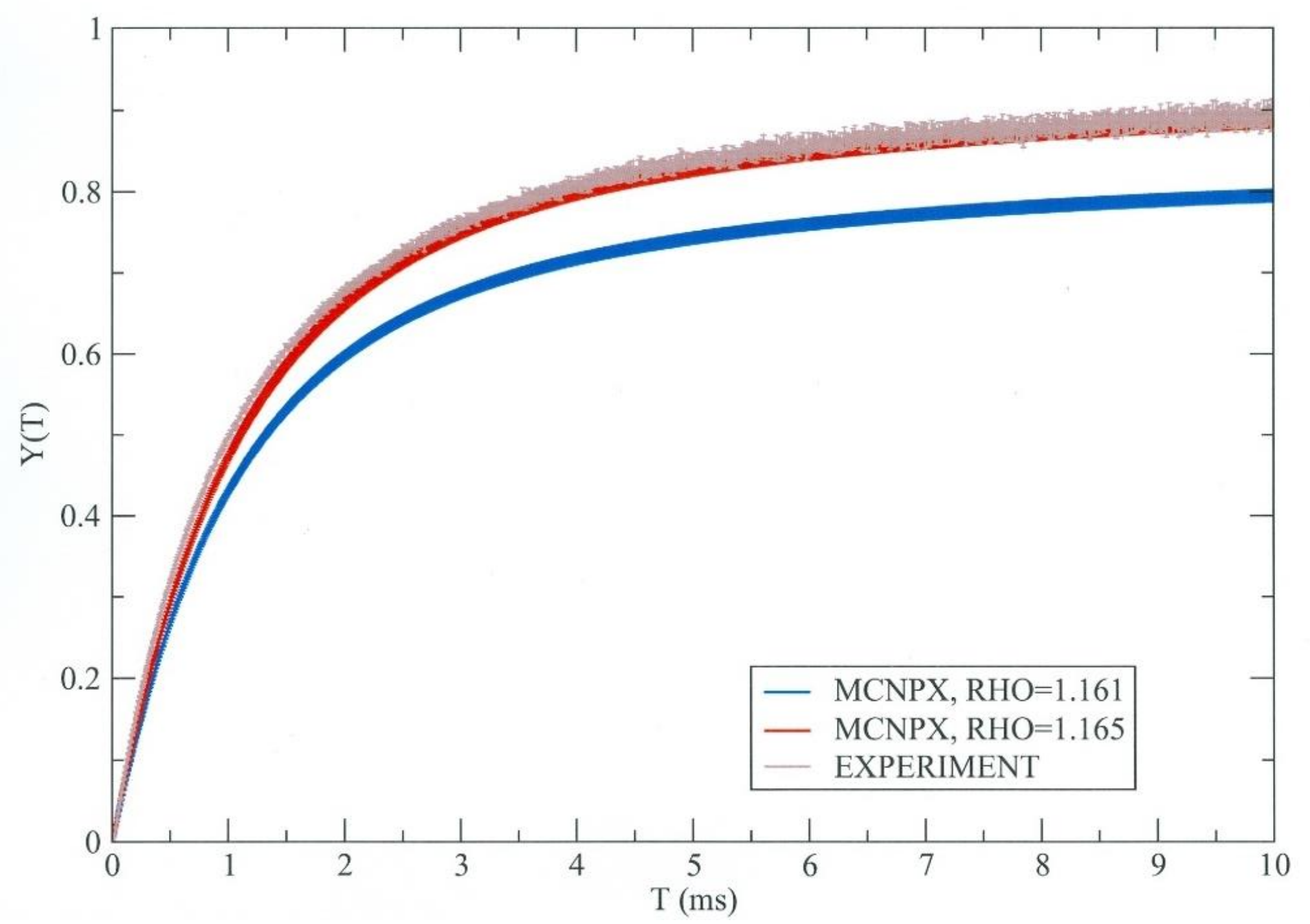

***Figure 12.*** Second order Feynman curves $Y(T)$. The blue curve is calculated with $\rho$ =1.161 g/cm$^3$, the red curve is calculated with $\rho$ =1.165 g/cm$^3$ and the measurements are in brown.

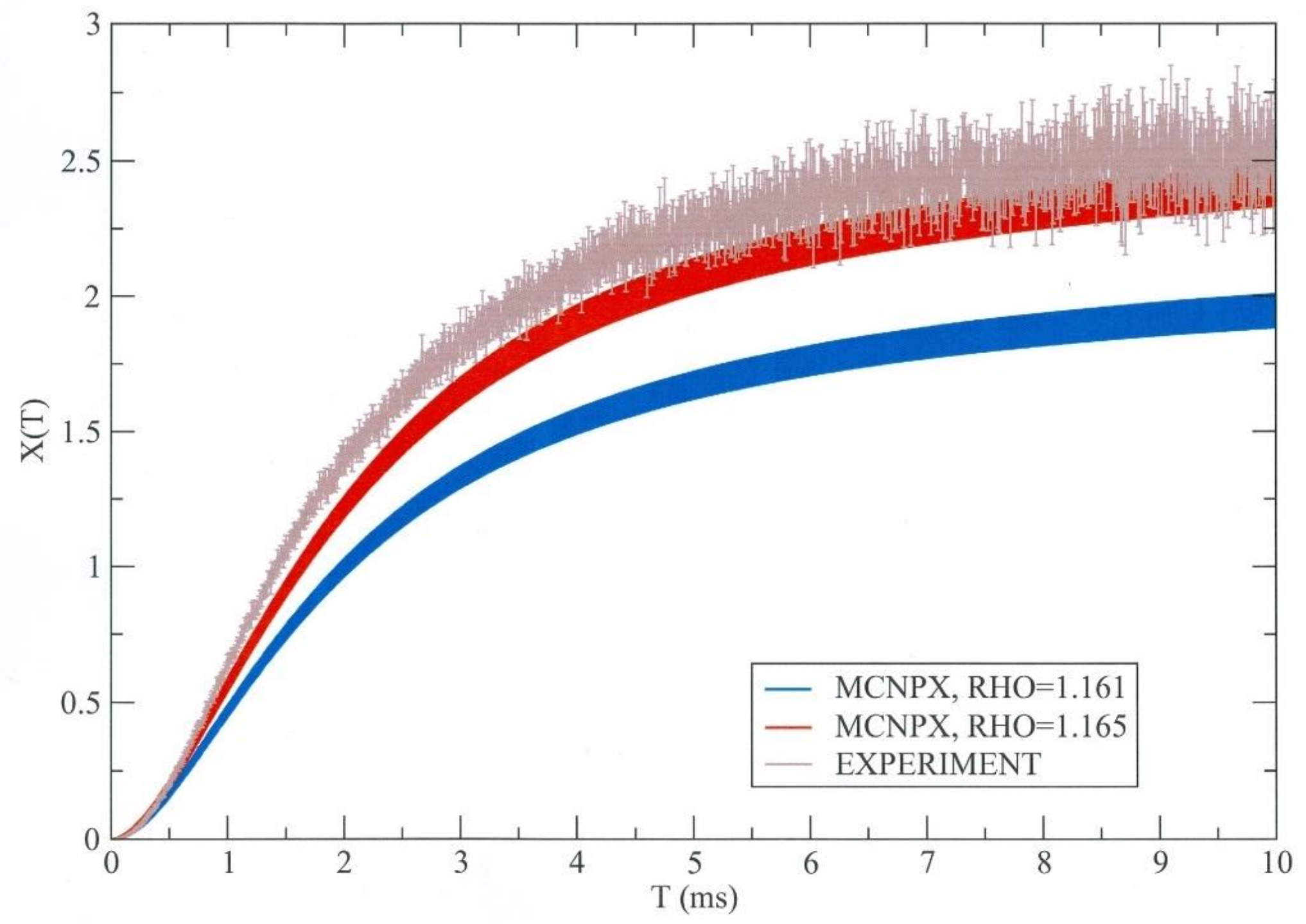

***Figure 13.*** Third order Feynman curves $Y(T)$. The blue curve is calculated with $\rho$ =1.161 g/cm$^3$. The red curve is calculated with $\rho$ =1.165 g/cm$^3$ and the measurements are in brown.

## VI. CONCLUSION

List mode measurements performed on SILENE reactor operated in stationary subcritical mode were analysed and simulated using the Feynman parameters method. The experimental time list file was processed using sequential binning in order to extract the second and third order Feynman factors. The Feynman parameters were obtained by non-linear least square regression and the uncertainties were evaluated with a non-overlapping block bootstrap method applied on the time list file.

The experiment was simulated using MCNPX code in analogue mode. The MCNPX-PTRAC coincidence capture file was used to generate a calculated time list file. This synthetic time list file contains the detection time intervals and the corresponding source event number. The file was processed using triggered binning taking into account only the detections correlated to the trigger signal. As for measurements, the calculated Feynman parameters were obtained using the least square regression and bootstrap.

Concerning the $Y_\infty$ and $X_\infty$ parameters, a better agreement between measured and calculated parameters was found using a fissile solution density a little higher than the nominal density. However the corresponding reactivity difference is small ($239.10^{-5}$).

## REFERENCES

1. R.E. UHRIG, "Random Noise Techniques in Nuclear Reactor Systems", the Ronald press company, (1970).
2. M.M.R. WILLIAMS, "Random Processes in Nuclear Reactors", Pergamon Press, (1974); https://doi.org/10.1016/C2013-0-05660-5
3. I. PAZSIT, L. PAL, *Neutron Fluctuation, a Treatise on the Physics of Branching Processes*, Elsevier, Oxford, (2008); https://doi.org/10.1016/B978-0-08-045064-3.X5001-7
4. B. VERREY et al, "Comparison of Subcritical Measurements with Calculated Results", *Transactions of the American Nuclear Society*, **93**, pp270-271, (2005).
5. D.B. PELOWITZ et al, "MCNPX User's Manual, Version 2.7.0", LA-CP-11-00438, (2011).
6. L.G. EVANS et al, "A New MCNPX PTRAC Coincidence Capture File Capability: A tool for Detector Design", Waste Management Symposium, Phoenix, USA, (2011).
7. R.P. FEYNMAN, "Dispersion of the Neutron Emission in U-235 Fission", *Journal of Nuclear Energy*, **3**, pp 64-69, (1956); https://doi.org/10.1016/0891-3919(56)90042-0
8. A. FURUHASHI, A. IZUMI, "Third Moment of the Number of Neutrons Detected in Short Time Intervals", *Journal of Nuclear Science and Technology*, **5:2**, pp48-59, (1968). https://doi.org/10.1080/18811248.1968.9732402
9. K. BÖHNEL, "The Effect of Multiplication on the Quantitative Determination of Spontaneously Fissioning Isotope of Neutron Correlation Analysis", *Nuclear Science and Engineering*, **90**, pp75-82, (1985). https://doi.org/10.13182/NSE85-2
10. D.M. CIFARELLI, W. HAGE, "Models for a Three Parameters Analysis of Neutron Signal Correlation Measurements for Fissile Material Assay", *Nuclear Instruments and Methods in Physics Research*, **A251**, pp550-563, (1986). https://doi.org/10.1016/0168-9002(86)90651-0
11. Y. KITAMURA et al, "Absolute Measurement of the Subcriticality by Using the Third Moment of the Number of Neutrons Detected", Proceedings of the PHYSOR 2002 conference, October 7-10, 2002, Seoul, Korea, (2002).

12. J. VERBEKE, “Neutron Multiplicity Counting : Credible Regions for Reconstruction Parameters”, *Nuclear Science and Engineering*, **182:4**, pp481-501, (2016). https://doi.org/10.13182/NSE15-35
13. F. BARBRY et al, “Uses and Performances of the SILENE Reactor”, Proceedings of the sixth international conference on criticality safety, ICNC99, (1999).
14. W.H. PRESS et al, “Numerical Recipes, the Art of Scientific Computing”, Cambridge University Press, (2007).
15. B. EFRON, “Bootstrap Method, another Look at the Jacknife”, *Annals of Stat*istics, **7**, pp1-26, (1979); https://doi.org/10.1214/aos/1176344552
16. T. ENDO, “Statistical Error Estimation Using Bootstrap Method for the Feynman−α Method”, *Transactions of the American Nuclear Society*, **111**, pp1204-1207, (2014).
17. H.R. KÜNCH et al, “The Jacknife and the Bootstrap for General Stationary Observations”, *Annals of Statistics*, **17**, pp1217-1241, (1989); https://doi.org/10.1214/aos/1176347265
18. S. CROFT, D. HENZLOVA, D.K. HAUK, ”Extraction of Correlated Count Rates using Gate Generation Techniques: Part I Theory”, *Nuclear Instruments and methods in Physics Research A*, **A69**, pp152-158, (2012). https://doi.org/10.1016/j.nima.2012.06.011
19. D. HENZLOVA et al, ”Extraction of correlated Count Rates using Gate Generation Techniques: Part II Experiment”, *Nuclear Instruments and methods in Physics Research A*, **A69**, pp159-167 (2012); https://doi.org/10.1016/j.nima.2012.04.091
20. J.D. ORNDORF, “Prompt Neutron Period of Metal Critical Assemblies”, *Nuclear Science and Engineering*, **2**, pp450-460, (1957); https://doi.org/10.13182/NSE57-A25409
21. M. CHADWICK et al., “ENDF/B-VII.1 Nuclear Data for Science and Technology: Cross Sections, Covariances, Fission Product Yields and Decay Data,” *Nuclear Data Sheets*, **112**, 12, 2887–2996, (2011); http://dx.doi.org/10.1016/j.nds.2011.11.002
22. P. GRIVOT et al, "SILENE Pulsed Experiments, SILENE-HEU-SOL-STEP S1-300/S2-300, S3-300", (2006);
23. T. MILLER et al, “Neutron Activation Foil and Thermoluminescent Dosimeter to a Bare Pulse of the CEA Valduc SILENE Critical Assembly”, NEA/NSC/DOC(95)03/VIII, ALARM-TRAN-AIR-SHIELD-001, ORNL/TM-2015/462 (2016).